# Targeting Melanoma-Specific Tyrosinase: Cyclic Peptide Disrupts Actin Dynamics for Precision Apoptosis Induction

Ruoyang Zhao[1], Xiaowei Wang[2], Jiajia Hu[1,3], Qingqing Sun[4], Xinmin Zhao[4], Jun Guo[1], Feng Zhang[1,4,5], Min Wu[1,3,6,7]

[1] Wenzhou Institute, University of Chinese Academy of Sciences, Wenzhou 325001, China.

[2] Shandong Guoyuan Human Genetic Resource Bank Management Co., Ltd, National Human Genetic Resources Sharing Service Platform-Shandong Innovation Center, Jinan 250117, China.

[3] Department of Biotherapy, Cancer Center and State Key Laboratory of Biotherapy, West China Hospital, Sichuan University, and Tianfu Jincheng Laboratory, Chengdu, China.

[4] Terahertz Technology Innovation Research Institute, Terahertz Spectrum and Imaging Technology Cooperative Innovation Center, Shanghai Key Lab of Modern Optical System, School of Optical-Electrical and Computer Engineering, University of Shanghai for Science and Technology, Shanghai 200093, China.

[5] National Center for Translational Medicine, Shanghai Jiao Tong University, Shanghai 200240, China.

[6] Department of Neurology, Medical College of Wisconsin, Milwaukee, WI 53226, USA.

[7] Department of Medicine, Harvard Medical School, and Brigham and Women's Hospital, Boston, MA 02115, USA.

**Correspondence**: Jun Guo (guojun-nbm@wiucas.ac.cn) | Feng Zhang (fzhang@usst.edu.cn) | Min Wu (miw100@ucas.ac.cn)

Ruoyang Zhao, Xiaowei Wang, and Jiajia Hu. contributed equally to this work.

## ABSTRACT

Melanoma is an aggressive and highly metastatic cancer that exhibits stubborn resistance to conventional therapies, highlighting the need for novel treatments. Existing therapeutic strategies often suffer from systemic toxicity, poor efficacy and fast-gained drug resistance. In this study, we designed a cyclic peptide system (c-RGDKYQ) that takes the advantage of the overexpression of tyrosinase in melanoma cells to trigger enzyme-mediated oxidation and self-assembly. The assembled peptide nanostructures can selectively disrupt the actin cytoskeleton, impairing cancer cellular functions, e.g., motility, adhesion, and proliferation, ultimately leading to apoptosis. This approach does not rely on external drug payloads or complex delivery mechanisms. c-RGDKYQ exhibits high selectivity for melanoma cells, strongly suppressing tumor growth in a murine model with minimal systemic toxicity. Our findings illuminate that, through targeting tyrosinase, c-RGDKYQ may be an enzyme-responsive alternative to conventional treatments for melanoma.

# 1 | Introduction

Melanoma is one of the most aggressive and lethal forms of cancer, characterized by rapid progression, high metastatic potential, and significant resistance to conventional treatments, such as chemotherapy and immunotherapy[1,2]. Despite advances in targeted therapies and immune checkpoint inhibitors, their clinical success is often compromised by systemic toxicity, off-target effects, and the risk of drug resistance[3]. These daunting challenges highlight the urgent need for innovative therapeutic strategies.

A distinctive feature of melanoma cells is the overexpression of tyrosinase, a melanocyte-specific enzyme central to melanin biosynthesis[4]. Tyrosinase can catalyze the oxidation of tyrosine and its derivatives within melanoma cells, which makes the tyrosinase as a promising biomarker and activator for tumor-specific therapeutic systems[5]. Recent advancements have harnessed tyrosinase activity to develop enzyme-responsive systems, such as prodrugs, nanoparticles, and pigment-mediated therapies[6,7]. These strategies primarily rely on tyrosinase-mediated oxidation to activate therapeutic agents or induce localized nanoparticle aggregation within melanoma cells, thereby exerting cytotoxic effects through the generation of reactive oxygen species (ROS) or controlled drug release. However, the dependence on exogenous payloads or intricate delivery systems presents substantial challenges in terms of scalability and clinical translation, ultimately constraining their broader applicability.

Additionally, the actin cytoskeleton, a critical component of cellular architecture and function, has emerged as a compelling therapeutic target[8]. Actin dynamics regulates key processes, such as cell motility, proliferation, and adhesion, which are essential for tumor invasion and metastasis[9]. Disrupting cytoskeletal dynamics represents a promising approach to impair these functions in cancer cells[10]. However, direct targeting cytoskeletal structures with tumor specificity has proven challenging, with existing strategies often hindered by off-target toxicity and insufficient therapeutic impact[11].

The discovery of the Arg-Gly-Asp (RGD) tripeptide sequence in the early 1980s marked a pivotal breakthrough in understanding cell-extracellular matrix (ECM) interactions. The RGD sequence was found to serve as a universal recognition site for integrin receptors - a family of transmembrane proteins critical for cellular processes such as adhesion, migration, and signaling[12,13]. synthetic RGD peptides could competitively inhibit cell attachment to fibronectin, demonstrating the sequence's essential role in integrin-mediated binding. Over the past four decades, RGD research has evolved from fundamental biochemical characterization to sophisticated biomedical applications[14]. Structural analyses elucidated that RGD-containing peptides adopt distinct conformations (cyclic vs. linear) that modulate integrin binding affinity and specificity. Notably, cyclic RGD variants demonstrated enhanced receptor selectivity, driving their development as anti-angiogenic agents in cancer therapy[15,16]. Parallel advances in nanotechnology and biomaterials leveraged the bio-adhesive property of RGD to engineer functionalized scaffolds for tissue regeneration and targeted drug delivery systems[15,17].

Tyrosinase catalyzes the initial and rate-limiting steps of melanin synthesis, converting tyrosine to DOPA and subsequently to dopaquinoine. Tyrosinase is highly expressed in melanoma cells, it plays a significant role in melanoma pathogenesis[4]. This overexpression provides a unique opportunity for targeted therapies, as normal cells exhibit minimal tyrosinase activity, reducing off-target effects. We have designed a series of sequences involving tyrosine (Tyr) and lysine (Lys), which are crucial for controlling self-assembly and functionality. The mechanism of action, driven by tyrosinase-catalyzed oxidation and sequence-guided self-assembly, enables the formation of fluorescent or melanin-like nanostructures[18-21]. By leveraging the unique overexpression of tyrosinase in melanoma cells, an unprecedent opportunity arises to selectively disrupt cytoskeletal dynamics through enzyme-triggered systems[22]. Such an approach should not only target key cellular processes but also address the limitations of current enzyme-responsive therapies by eliminating the need for external drug payloads and complex delivery mechanisms[23,24]. This strategy is promising for significantly enhancing therapeutic selectivity while minimizing systemic toxicity, providing a novel framework for melanoma treatment.

In this study, we introduce a novel cyclic peptide system (c-RGDKYQ) that leverages the overexpression of tyrosinase in melanoma cells to achieve in situ self-assembly. The cyclic peptide is synthesized through an ortho-phthalaldehyde-mediated cyclization reaction under mild aqueous conditions. Upon tyrosinase-catalyzed oxidation, c-RGDKYQ generates reactive quinone structures, which drives the formation of supramolecular nanostructures (Figure 1A). These nanostructures selectively disrupt the actin cytoskeleton dynamics within melanoma cells, impairing cellular motility, proliferation, and adhesion, and ultimately inducing apoptosis (Figure 1B). Notably, this approach simplifies the therapeutic design and facilitates its translational potential. Both *in vitro* and *in vivo* experimental results validate the high selectivity of c-RGDKYQ for melanoma cells, demonstrating significant tumor growth suppression with minimal adverse effects. These findings, collectively, establish c-RGDKYQ as a robust and efficient framework for enzyme-responsive cancer therapies, specifically targeting melanoma and other tyrosinase-expressing tumors.

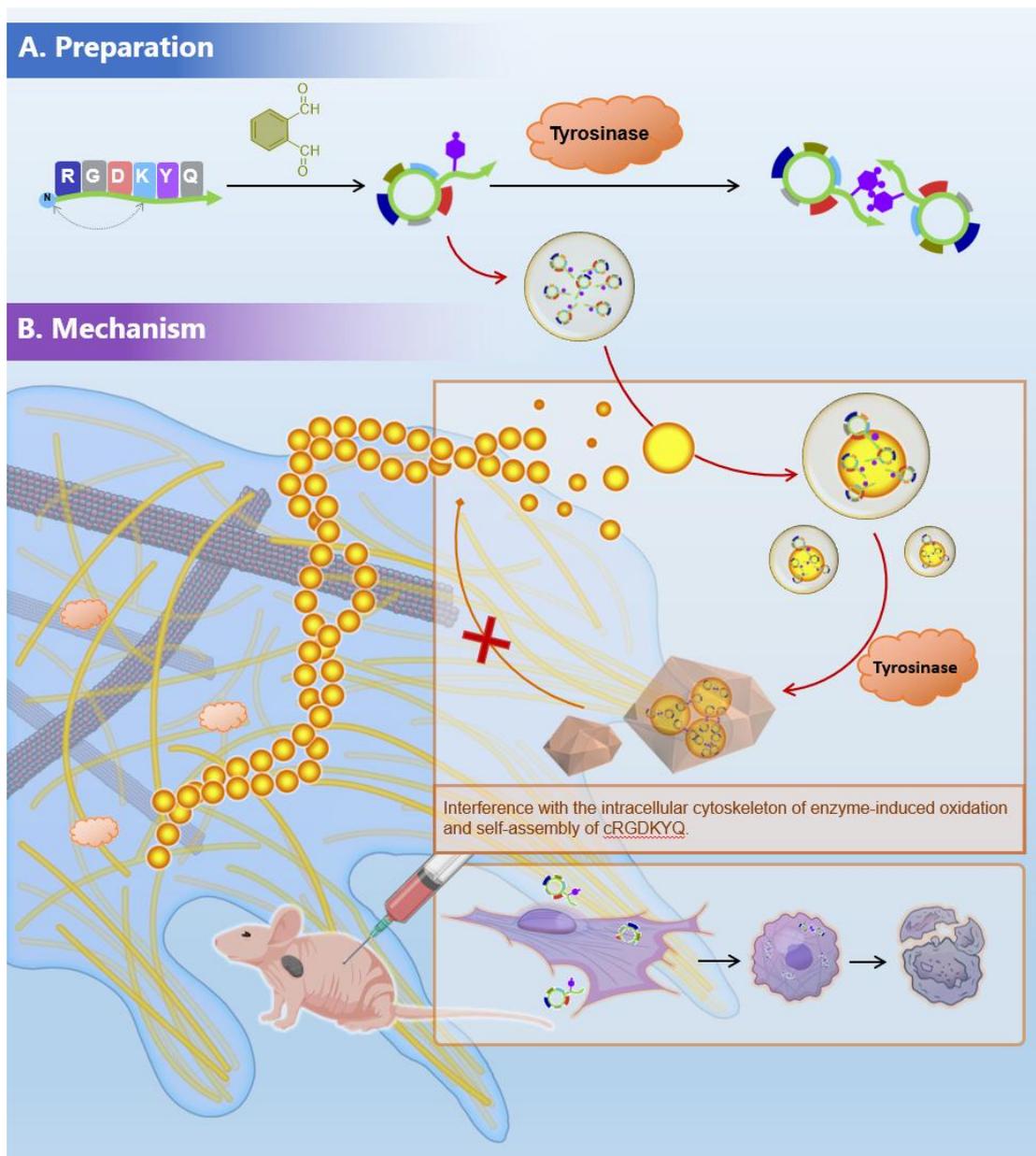

**FIGURE 1 | Tyrosinase-induced cyclic peptide self-assembly and its anti-melanoma mechanism. (A)** Synthesis of the cyclic peptide: The peptide chain containing tyrosine and lysine residues undergoes cyclization through an ortho-phthalaldehyde-mediated reaction, forming the cyclic peptide. Upon tyrosinase catalysis, the cyclic peptide undergoes tyrosine oxidation, generating quinone groups which prefer to self-assemble into nanostructures. **(B)** Molecular mechanism: The cyclic peptide disrupts the polymerization/fibrillization of actin proteins to form the cytoskeleton, thereby inhibiting the growth of melanoma cells. Following peri-tumoral injection, the cyclic peptide significantly suppresses the growth of drug-resistant melanoma cells by activating apoptosis pathways.

## 2 | Results

### 2.1 | Peptide Cyclization, Oxidation and Self-Assembly

The cyclic peptide (c-RGDKYQ) was successfully synthesized using an ortho-phthalaldehyde (OPA)-mediated cyclization reaction (Figure 2A). First, we tested whether this peptide can be selectively activated by tyrosinase via undergoing oxidation and further self-assembly. After exposure to tyrosinase, the peptide solution color changed from transparent to deep yellow, indicating the formation of oxidative products and nanostructures (Figure 2B). UV-vis spectroscopy revealed characteristic absorption peaks between 330 and 450 nm, corresponding to the formation of aromatic and quinone structures through tyrosinase-mediated oxidation of tyrosine residues (Figure 2C). Fluorescence spectroscopy showed a strong emission peak at 450 nm when excited at 340 nm, confirming the structural transformation necessary for self-assembly (Figure 2D).

Peptide solution incubated with tyrosinase self-assembled into nanostructures with a narrow size distribution, which was much larger and more uniform compared to those of untreated peptide aggregates (Figure 2E). This transition was accompanied by changes in $\zeta$-potential, which reflected the increased hydrophobicity and stability of the formed nanostructures (Figure 2F). HPLC analysis further confirmed the enzyme-specificity of the process, as oxidation and nanostructure formation did not occur in control samples lacking tyrosinase (Figure. S1). FTIR indicates that the cyclization involving aromatic groups causes a change in the absorption of C-H out-of-plane bending vibrations in the benzene ring at 925 $cm^{-1}$. Both linear and cyclic peptides, under the catalysis of tyrosinase, exhibit oxidation and polymerization of the phenolic hydroxyl group on the tyrosine amine, as evidenced by the reduction in characteristic FTIR peaks between 1138 and 1208 $cm^{-1}$ and an increase of 1400 $cm^{-1}$ (Figure 2G). The tyrosinase-mediated oxidation of tyrosine residues to reactive quinones could promote intermolecular crosslinking and the formation of melanin-like polymers that subsequently assemble into nanostructures (Figure 2H). This targeted transformation underscores the unique capability of the cyclic peptide to exploit the enzymatic environment of melanoma cells for selective activation and assembly.

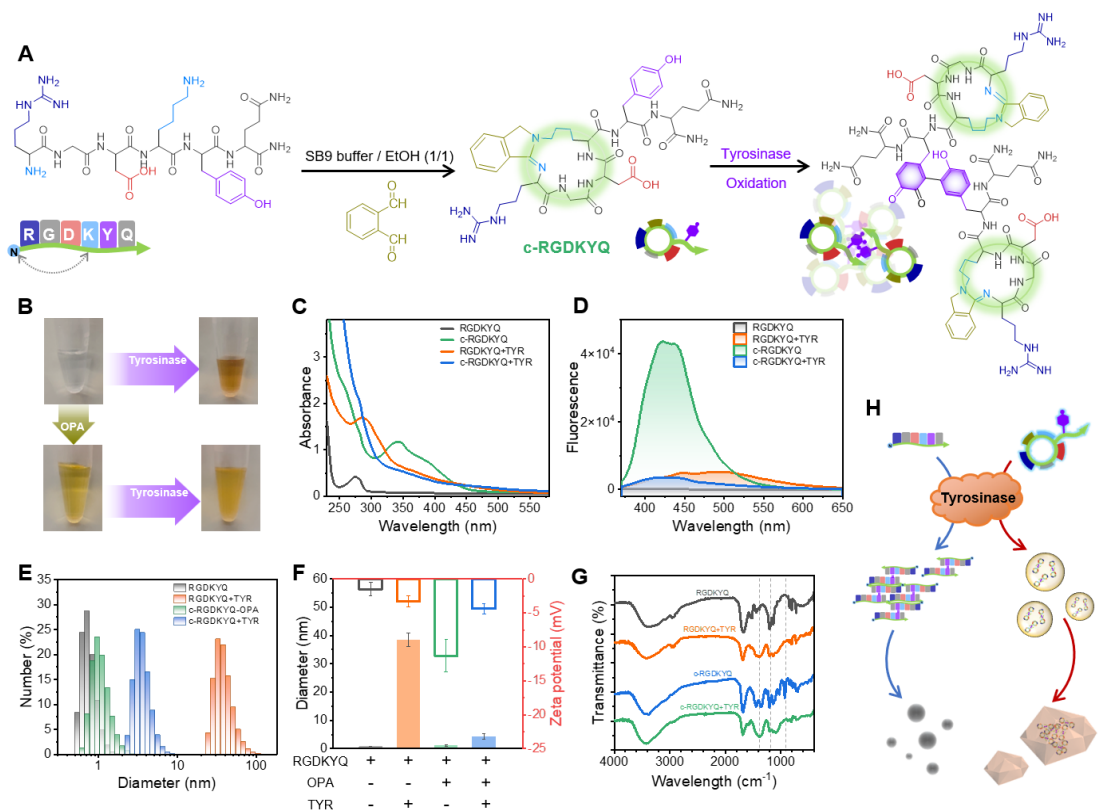

**FIGURE 2 | Peptide cyclization, oxidation and self-assembly.** (A) Schematic illustration of the synthesis of cyclic peptide (c-RGDKYQ) via ortho-phthalaldehyde (OPA)-mediated condensation under aqueous conditions and subsequent tyrosinase-catalyzed oxidation and crosslinking. (B) Color changes during the cyclization and enzymatic reaction. The uncyclized peptide solution (transparent) turned light yellow after OPA treatment and dark yellow upon the addition of a tyrosinase, indicating tyrosine oxidation and self-assembly. (C) UV-vis absorption spectra revealed new peaks at 330–450 nm during the cyclization and oxidation processes. (D) Fluorescence spectra of the cyclic peptide after oxidation, exhibiting a strong emission peak at 450 nm when excited at 340 nm, confirming the presence of aromatic structures. (E) Hydrodynamic size comparison of peptide species by dynamic light scattering (DLS) measurements. (F) Average hydrodynamic size and zeta potential of the cyclic peptide under different conditions. The data are presented as mean ± standard deviation (s. d.) from three independent experiments. (G) FTIR spectra of RGDKYQ, c-RGDKYQ and RGDKYQ + tyrosinase (TYR) and c-RGDKYQ + TYR. (H) A schematic representation of the cyclization

and self-assembly process illustrated the transformation from peptide monomers to oxidized melanin-like polymers and nanostructures.

**2.2 | Selective Cytotoxicity to Melanoma Cells**

The cytotoxic effects of the linear peptide (RGDKYQ) and its oxidized form (c-RGDKYQ) were systematically evaluated across melanoma cells (B16) and two other cell lines (L929 and U2OS). RGDKYQ exhibited negligible cytotoxicity at all tested concentrations in all cell lines. In contrast, c-RGDKYQ demonstrated a concentration-dependent cytotoxic effect, significantly reducing cell viability in B16 melanoma cells but sparing the L929 and U2OS cells (Figure 3A-B). This selective cytotoxicity underscores the tyrosinase-mediated activation of the peptide within melanoma cells, which overexpress the enzyme. Morphological observations of B16 cells treated with c-RGDKYQ revealed dose-dependent cellular damage, including disruption of the cellular architecture and increased cell death (Figure 3E, S2). These findings align with the role of the oxidized cyclic peptide in impairing cytoskeletal integrity, a critical aspect of cellular survival and adhesion. Apoptosis analysis demonstrated that with the increase of c-RGDKYQ concentration, the number of apoptosis increased significantly (Figure 3F, G).

To assess the impact of c-RGDKYQ on cell motility, a wound-healing assay was performed on B16 melanoma cells. Untreated cells and cells exposed to RGDKYQ showed progressive wound closure over 24 hours. However, treatment with c-RGDKYQ effectively inhibited cell migration, as evidenced by the delayed wound closure compared to control groups (Figure 3H, I). These results demonstrate that the oxidized cyclic peptide disrupts cellular motility, further supporting its anti-melanoma activity. Collectively, these findings demonstrate the selective cytotoxicity and anti-motility effects of c-RGDKYQ in melanoma cells, highlighting its potential for targeted melanoma therapy, while minimizing off-target effects in other cells.

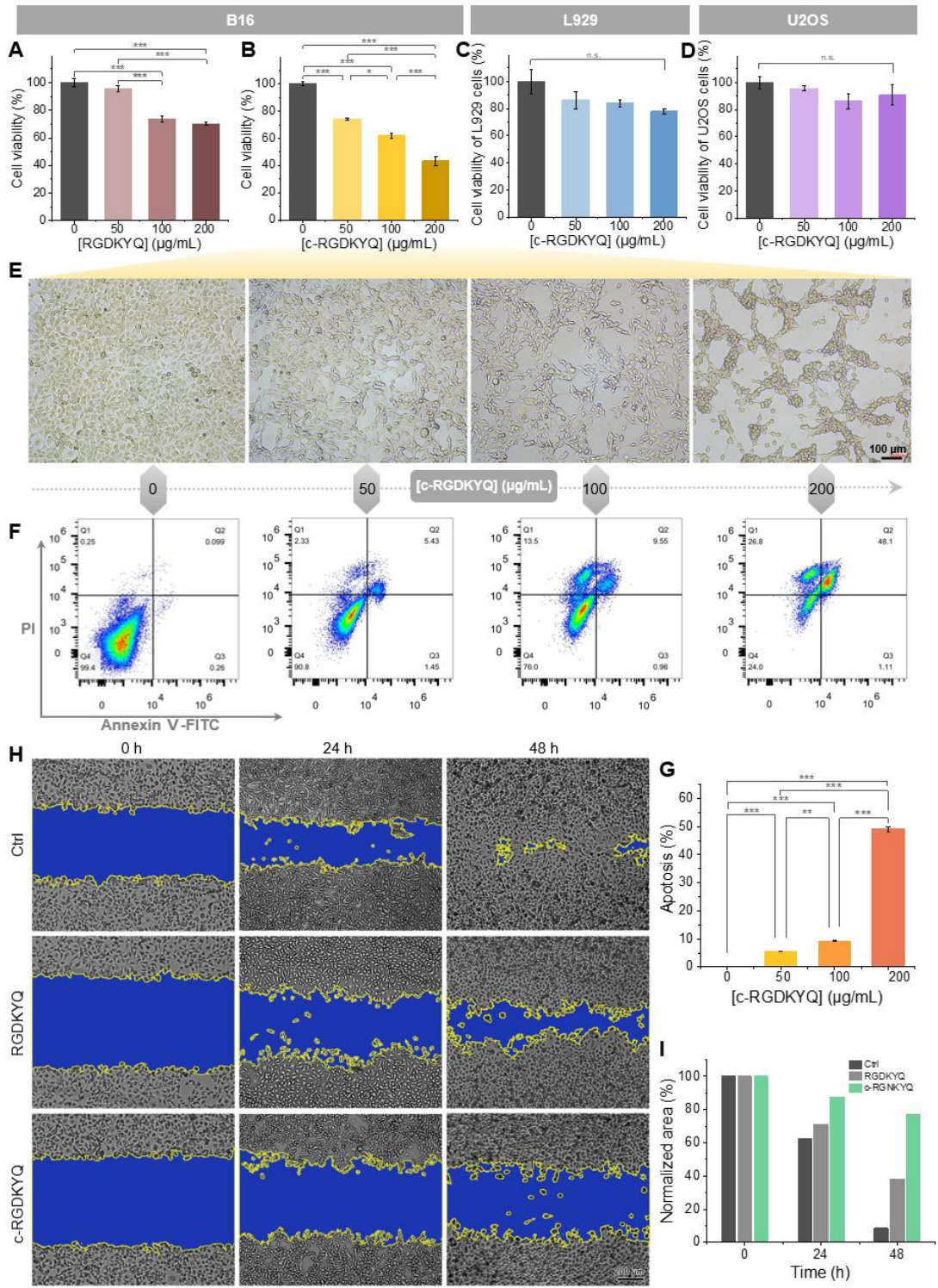

**FIGURE 3 | Selective cytotoxicity and inhibition of cell migration.** (A-D) Effects of cyclic peptide on cell viability across different cell lines. (A) RGDKYQ in the viability of B16 cells. (B) c-RGDKYQ exhibited enhanced cytotoxicity in B16 cells. Viability of L929 (C) and U2OS

(D) cells treated with c-RGDKYQ. (E) Morphological changes in B16 cells upon treatment with increasing concentrations of c-RGDKYQ (0, 50, 100, 200 µg/mL), revealing cell aggregation and structural damage. (F) Representative Annexin V-FITC/PI co-staining flow cytometry results of B16 cells after incubation with 0, 50, 100, and 200 µg /mL c-RGDKYQ for 24 h. (G) Statistics of apoptosis (Annexin V-FITC/PI co-staining) of B16 cells. The experiments were repeated for three times and data was presented as mean ± s. d. (H) Cell migration was determined by a wound-healing assay of B16 cell monolayers. Untreated control (Ctrl) cells showed substantial migration closure at 24 h and 48 h, whereas migration was notably inhibited in RGDKYQ and c-RGDKYQ-treated B16 cells. (I) Quantification of migration gaps. Statistical analysis: all experiments were repeated three times, with data presented as mean ± standard deviation (mean ± s. d.). One-way ANOVA was used for statistical analysis. Significant levels are indicated as follows: n. s., not significant; *$p ≤ 0.05$; **$p ≤ 0.01$; ***$p ≤ 0.001$.

**2.3 | Dynamic Disruption of Actin Cytoskeleton**

The effects of c-RGDKYQ on the actin cytoskeleton of B16 melanoma cells were investigated using fluorescence microscopy. In Ctrl group, phalloidin staining revealed intact actin filaments organized in dense, fibrous networks, reflecting normal cytoskeletal integrity (Figure 4A, B). In contrast, B16 cells treated with c-RGDKYQ for 24 hours displayed a striking disruption of actin filaments. The fluorescence intensity of phalloidin-stained filaments significantly decreased, and the structural organization of actin filaments appeared fragmented and disordered (Figure 4C, D). This disruption correlated with impaired cell spreading and adhesion, suggesting that c-RGDKYQ interferes with the dynamic remodeling of the actin cytoskeleton. The observed effects can be attributed to the selective oxidation of peptide by tyrosinase, which generates reactive quinones that inhibit actin polymerization. A schematic representation of this mechanism (Figure 4E) illustrates how c-RGDKYQ disrupts cytoskeletal growth, leading to loss of cellular structural integrity and inducing apoptosis in melanoma cells. The severe reduction in actin filament organization is consistent with morphological abnormalities and loss of cellular adhesion observed in previous experiments. The above results show that c-

RGDKYQ can specifically target cytoskeletal components in melanoma cells, which takes the advantage of the high levels of tyrosinase expression in these tumor cells. By disrupting actin dynamics, c-RGDKYQ impairs critical cellular processes, such as motility and structural maintenance, contributing to its selective cytotoxic effects, indicating the therapeutic potential of targeting cytoskeletal dynamics in melanoma treatment.

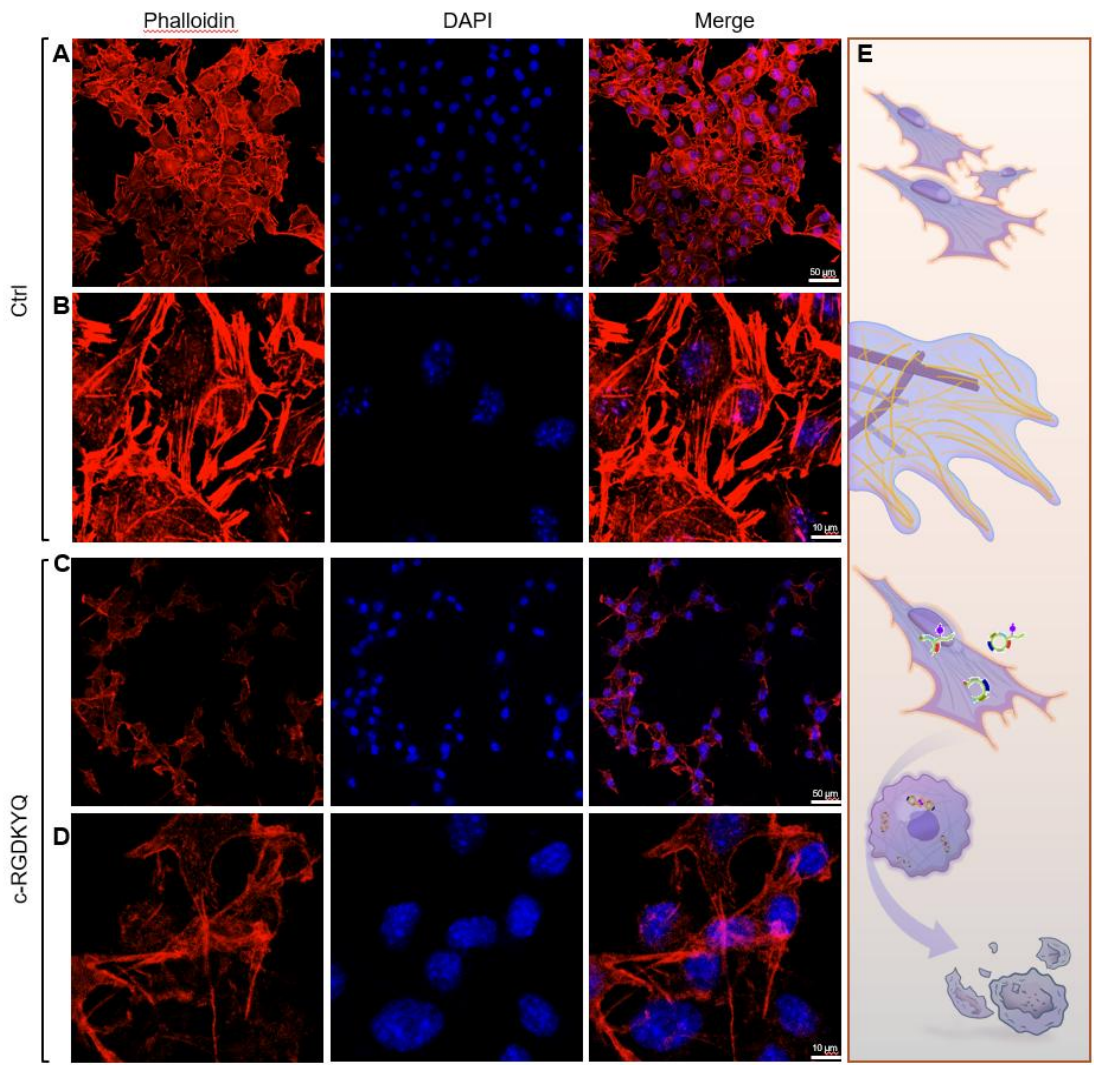

**FIGURE 4 | Disrupting cytoskeleton in melanoma cells.** (A-B) Phalloidin staining (red) in B16 cells (Ctrl) showed intact actin filaments organized in a typical fibrous pattern, with nuclei counterstained by DAPI (blue). The cytoskeleton structure was well-preserved. (C-D) B16 cells treated with c-RGDKYQ for 24 hours displayed significant reduction in red fluorescence,

showing disrupted actin filaments and impaired cytoskeletal organization. (E) A schematic mechanism showing that oxidation and assembly of the peptide catalyzed by tyrosinase, further interferes with actin polymerization, disrupts cytoskeletal dynamics, and eventually leads to apoptosis.

**2.4 | Therapeutic Efficacy in a Murine Melanoma Model**

The therapeutic efficacy of c-RGDKYQ was assessed in a murine melanoma model using B16 melanoma cells grafted into the right flank of mice. The mice were treated with peri-tumoral injections of saline (Ctrl), RGDKYQ, or c-RGDKYQ according to the schedule shown in Figure 5A. Body weight measurements throughout the study showed no significant differences between treatment groups, suggesting minimal systemic toxicity of c-RGDKYQ (Figure 5B). We next found that tumor growth was markedly inhibited in the c-RGDKYQ-treated group compared to both the Ctrl and RGDKYQ-treated groups. Tumor volume measurements over time revealed pronounced suppression of tumor growth in the c-RGDKYQ group (Figure 5C). The final tumor weights further confirmed the efficacy of c-RGDKYQ, with treated mice exhibiting markedly smaller tumor masses than Ctrl (Figure 5D). Representative photographs of tumor-bearing mice and excised tumors illustrate the substantial tumor reduction achieved by c-RGDKYQ treatment (Figure 5E, F). Histological examination of tumor sections stained with H&E revealed significant differences in tumor architecture. c-RGDKYQ-treated tumors showed increased apoptotic regions and reduced cellular density compared to control and RGDKYQ-treated tumors (Figure 5G), consistent with the *in vitro* experimental results, in which c-RGDKYQ disrupted actin cytoskeleton dynamics and induced apoptosis. Collectively, we demonstrated the strong antitumor efficacy of c-RGDKYQ *in vivo*, selectively targeting melanoma cells without causing systemic toxicity. The select activation of the cyclic peptide by tyrosinase and its subsequent therapeutic effects emphasize its potential as a highly specific and effective treatment for melanoma.

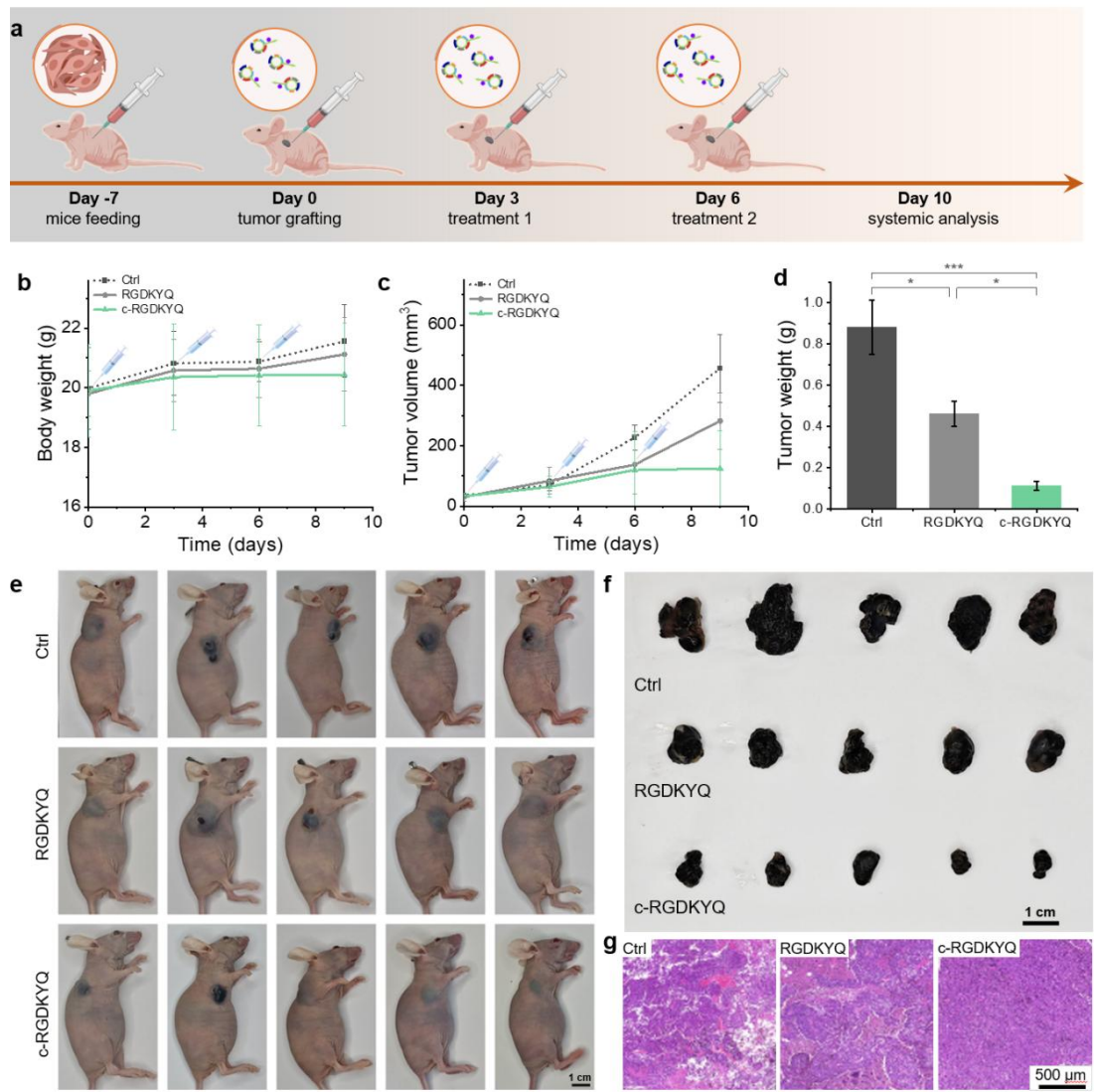

**FIGURE 5 | Antitumor activity of cyclic peptide in a murine melanoma model.** (A) Schematic timeline of the melanoma model establishment and treatment regimen. On day 0, B16 cells were subcutaneously grafted into the right flank of mice. Tumor volume was monitored until it reached ≥30 mm³, after which peritumoral injections were initiated on days 3 and 6. Systemic analysis was conducted on day 10. (B) Body weight curves of mice in different treatment groups (Ctrl: saline; RGDKYQ: uncyclized peptide; c-RGDKYQ: cyclic peptide). (C) Tumor volume growth curves. (D) Statistical analysis of tumor weights after treatment. (*p ≤ 0.05, **p ≤ 0.01, ***p ≤ 0.001). e Photographs of mice from different treatment groups showing markedly smaller tumor sizes in the cyclic peptide-treated group. (F) *Ex vivo*

tumor images illustrating the differences in tumor size and weight among treatment groups. (G) H&E-stained tumor sections. Tumor tissues from the cyclic peptide-treated group exhibited increased cellular apoptosis and necrosis, evidenced by morphological changes.

## 3 | Discussion

The designed cyclic peptide can leverage the overexpression of tyrosinase in melanoma cells to achieve selective activation and therapeutic effects. Through tyrosinase-mediated oxidation, the cyclic peptide undergoes a transformation into reactive quinone structures, leading to the formation of nanostructures that selectively disrupt melanoma cell actin cytoskeleton dynamics. This targeted disruption impairs critical cellular functions, such as motility, adhesion, and proliferation, and ultimately inducing apoptosis. Our results underscore the effectiveness of c-RGDKYQ for addressing key challenges in melanoma therapy, including tumor specificity and the limitations of conventional drug delivery systems.

Research demonstrates that the RGD sequence regulates interactions of intrinsically disordered region (IDR)-containing proteins via integrin receptor binding, influencing cellular signaling, cytoskeletal reorganization, and phase separation. Cyclic RGD peptides exhibit enhanced binding affinity due to their stable conformation, activating FAK/Src/Rho GTPase pathways to strengthen cell-matrix adhesion and promote dynamic cytoskeletal remodeling (e.g., actin polymerization), significantly boosting cell migration efficiency. This mechanism is critical in tumor invasion and angiogenesis, where cyclic RGD peptides enhance cancer cell penetration through basement membranes. Additionally, cyclic RGD peptides are utilized in tumor-targeted therapies and drug delivery systems by modulating cytoskeletal dynamics and phase separation within the tumor microenvironment. Functional stability and therapeutic efficacy are further optimized through aromatic group modifications during cyclization.

Here *in vitro* experiments showed that c-RGDKYQ selectively inhibits the viability of B16 melanoma cells while sparing other cells (L929 and U2OS), highlighting its specificity for tyrosinase-expressing cells. This specificity was further validated through the analysis of cellular tyrosinase activity, which was significantly higher in B16 cells compared to the other cell lines. Importantly, the cyclic peptide not only demonstrated cytotoxic effects but also impaired cell motility, as evidenced by the wound-healing assays. The ability to simultaneously disrupt actin dynamics and impair migration suggests a potentially multi-faceted mechanism for suppressing melanoma progression. The *in vivo* evaluation of c-RGDKYQ in a murine melanoma model further supported its therapeutic potential. Tumor growth was significantly suppressed in the c-RGDKYQ-treated group, with treated mice showing markedly smaller tumor volumes and weights compared to the control and the group treated with the unoxidized peptide. The histological analysis revealed increased apoptotic regions and reduced cellularity in tumors from the c-RGDKYQ group, indicating effective induction of cell death within the tumor microenvironment. Critically, the treatment did not cause significant changes in body weight, suggesting minimal systemic toxicity.

A key advantage of c-RGDKYQ system is its simplicity/feasibility and lack of reliance on external drug payloads or complex delivery vehicles. The cyclic peptide itself serves as both the therapeutic agent and the responsive system, which significantly reduces formulation complexity and enhances translational potential. Unlike conventional approaches that depend on nanoparticle carriers or prodrug formulations, c-RGDKYQ offers a streamlined therapeutic strategy with high tumor selectivity and minimal off-target effects. The selective disruption of the actin cytoskeleton, driven by the intracellular self-assembly of the oxidized peptide, is a particularly promising aspect of this work. By targeting cytoskeletal dynamics, c-RGDKYQ can directly interfere with processes essential for tumor cell survival and invasion, which aligns with emerging evidence highlighting the cytoskeleton as a critical therapeutic target in cancer. Moreover, targeting tyrosinase as a tumor-specific enzymatic trigger provides a robust platform for further development of enzyme-responsive therapeutic systems.

## 4 | Conclusion

In summary, c-RGDKYQ is successfully designed as a highly selective and potent therapeutic agent for melanoma, capitalizing on the overexpression of tyrosinase in melanoma cells. Through enzyme-triggered oxidation and subsequent in situ self-assembly, the cyclic peptide selectively disrupts actin cytoskeleton dynamics, leading to cancer cell apoptosis. The demonstrated efficacy in both *in vitro* and *in vivo* experiments underscore its potential as a targeted, tumor-specific treatment strategy. Notably, the feasibility, specificity, and minimal systemic toxicity represent the translational potential of c-RGDKYQ for melanoma therapy. Future investigations will explore its applicability to other tyrosinase-expressing malignancies and further optimize delivery and efficacy in clinical settings.

## 5 | Materials and Methods

The peptide sequencing (RGDKYQ) was custom-synthesized by Top Peptide Biotechnology Co. Ltd. (Shanghai, China). Tyrosinase ($\geq$ 500 units/mg protein) and o-phthalaldehyde (OPA, molecular biology grade) were supplied by Sigma-Aldrich and purchased through Merck China Co., Ltd. All materials were used as received without further purification. Water was purified using a Millipore system (Sigma Aldrich, American) with a minimum resistivity of 18.2 M$\Omega$·cm.

### 5.1 | Preparation of c-RGDKYQ

Unprotected peptide RGDKYQ were dissolved in phosphate buffer (pH 8.0) saline /EtOH (1:1) to give a clear solution. OPA (1.0 equiv, stock solution) was added to the solution at room temperature and stirred for 10 min. The freeze-dried product is a light-yellow powder.

### 5.2 | Absorbance and fluorescence spectroscopy

Fluorescence spectroscopy was performed on a fluorescence spectrometer (Fluorolog®-MAX 4, Horiba) equipped with a 1.0 cm quartz cuvette with a fixed excitation wavelength at 350 nm. Both excitation and emission slit widths were set 5.0 nm. The absorbance spectroscopy was performed on a microplate reader (BioTek Synergy H1).

### 5.3 | Infrared spectrometry

Infrared (IR) spectra were recorded on Nicolet iS5 FTIR spectrophotometer (Thermo Scientific) with KBr pellets in the 4000 - 400 cm$^{-1}$ regions.

### 5.4 | Size and zeta potential

Diameter and zeta potential measurements were performed using a dynamic light scattering (DLS) instrument, specifically the Malvern Zetasizer Nano ZS90 (Malvern Panalytical Ltd.), equipped with a quartz cuvette. The experiments were conducted at a controlled temperature of

25.0 °C over 100 consecutive scans. The samples were prepared in a sodium borate buffer (10 mM, pH adjusted between 3 and 10) to ensure optimal conditions for accurate measurements.

**5.5 | High performance liquid chromatography and mass spectrometry**

High performance liquid chromatography (HPLC) analyses were conducted using an Agilent Technologies 1260 Infinity system, fitted with an Inertsil ODS-SP column (GL Sciences Inc., 5 μm particle size, 4.6 mm × 250 mm dimensions). The mobile phase consisted of a sodium borate buffer (10 mM, pH 7). Detection was performed using a UV detector set at 280 nm and a fluorescence detector configured with an excitation wavelength of 350 nm and an emission wavelength of 500 nm.

Following HPLC separation, mass spectrometry (MS) analysis was carried out to identify and quantify the compounds. The HPLC system was coupled online with a mass spectrometer (Agilent 6460 Triple Quadrupole LC/MS or equivalent), which operated in both positive and negative ion modes to maximize detection sensitivity. The electrospray ionization (ESI) source parameters were optimized for each sample type, including capillary voltage, nebulizer gas pressure, drying gas flow rate, and temperature. Data acquisition and processing were managed using MassHunter Workstation Software (Agilent Technologies). The MS conditions were tailored to ensure accurate mass measurement and reliable compound identification, facilitating comprehensive characterization of the samples.

**5.6 | Cell proliferation and viability assay**

CCK-8 assay kit (K1080, Shanghai WeiHuan Biotech CO., Ltd.) was utilized to assess cell proliferation and viability. Log-phase B16, L929, and U2OS cells were dissociated using 0.25% trypsin to form single-cell suspensions. Cells were counted using a hemocytometer and adjusted to a concentration of $5 \times 10^4$ cells/mL in complete medium. A volume of 100 μL of the cell suspension was seeded into each well of a 96-well plate, with three replicates per condition. The plates were incubated at 37 °C with 5% $CO_2$ for 24 hours to allow for cell attachment. Following incubation, different concentrations of the test drug (three concentration groups) were added to the wells. Each treatment group had three replicates. After 12 hours, images were captured under a microscope to document initial cell viability. For B16 cells, the 96-well plates were further incubated at 37 °C with 5% $CO_2$ for an additional 24 hours. Subsequently, 10 μL of CCK-8 reagent was added to each well, gently mixed, and the plates were incubated for 2.5 hours under the same condition. Absorbance values were measured at 450 nm using a microplate reader (U-2900, Hitachi). Data were recorded, and the percentage of cell viability was calculated. Origin software was used for data analysis to generate cell viability curves and determine the half-maximal inhibitory concentration.

### 5.7 | Wound healing assay for cell migration

To evaluate the effects of Control, RGDKYQ, and cRGDKYQ on cell migration, a classical wound healing assay was performed. B16 cells in logarithmic growth phase were seeded at a density of $2 \times 10^5$ cells per well in 6-well plates and cultured in DMEM medium supplemented with 10% fetal bovine serum (FBS) and 1% penicillin-streptomycin at 37 °C with 5% $CO_2$ until they reached 90%-100% confluency. Cells were then starved in serum-free medium for 24 hours to synchronize the cell cycle and minimize non-specific migration. Next, a straight line was gently scratched in the center of each well using a 200 μL pipette tip, creating a uniform "cell-free zone," followed by twice washes with PBS to remove unattached cell debris. For treatment, the control group received normal culture medium, the RGDKYQ group was treated with a final concentration of 0.1 mg/mL, and the cRGDKYQ group received a final concentration of 0.1 mg/mL. Images of the scratch areas in the center of each well were captured at 0, 12, and 24 hours using an inverted microscope (OLYMPUS DP74), ensuring consistent imaging positions. The changes in scratch width at different time points were measured using ImageJ software to calculate relative migration distances or percent closure of the scratch. Statistical analysis was performed using T-tests or ANOVA to compare differences between the three groups, with $p < 0.05$ considered statistically significant.

### 5.8 | Flow cytometry assay for apoptosis analysis

Flow cytometry was employed to analyze cell cycle progression and apoptosis. B16 cells in the logarithmic growth phase were seeded in 6-well plates at a density of $2 \times 10^5$ cells per well. After reaching 70–80% confluency, cells were synchronized by serum starvation for 24 hours. Subsequently, the cells were treated with Control and cRGDKYQ (0 mg/mL, 50 mg/mL, 100 mg/mL and 200 mg/mL) for 24 hours. The Apoptosis Analysis was performed using Annexin V-FITC/PI Apoptosis Detection Kit (Catalog No. K2003, Weihuan Biotechnology Co., Ltd.) following the manufacturer's protocol. Stained cells were diluted with 500 μL buffer and analyzed immediately on a CytoFlex Flow Cytometer (Beckman Coulter). A minimum of

50,000 events per sample were acquired, and data were processed with CytExpert Software (v2.4). Statistical significance was determined by one-way ANOVA with T-test ($p < 0.05$; n = 3 independent experiments, triplicate wells).

**5.9 | Observation and analysis of cytoskeletal structure**

Cells were cultured in 24-well plates and until reaching approximately 70-80% confluency. The cells were maintained in DMEM medium supplemented with 10% FBS and 1% penicillin-streptomycin at 37 °C in a 5% $CO_2$ incubator. Cells were gently washed twice with pre-cooled PBS, each for 5 minutes. Fixation was performed by adding 4% paraformaldehyde (PFA) solution and incubating at room temperature for 15 minutes. Following fixation, cells were washed twice with pre-cooled PBS, each for 5 minutes. To allow the dye to penetrate the cell membrane, cells were permeabilized with 0.1% Triton X-100 solution at room temperature for 5 minutes. Non-specific binding sites were blocked by incubating the cells with PBS containing 1% bovine serum albumin (BSA) for 30 minutes at room temperature. Cells were then stained with a PBS solution containing Alexa Fluor 488-labeled Phalloidin diluted 1:100 and incubated for 30 minutes in the dark. Phalloidin specifically binds to F-actin, labeling the actin filaments of the cytoskeleton. After staining, cells were washed three times with PBS, each for 5 minutes. For nuclear staining, cells were incubated with a PBS solution containing DAPI (4',6-diamidino-2-phenylindole) diluted 1:1000 for 10 minutes in the dark at room temperature. Cells were washed again three times with PBS, each for 5 minutes. Coverslips were carefully removed and inverted onto glass slides containing an anti-fade mounting medium, sealing the edges with nail polish. Fluorescence images were captured using a ZEISS Axio Observer A1 microscope equipped with appropriate filters. F-actin fluorescence was observed under excitation at 488 nm, while DAPI fluorescence was detected under excitation at 358 nm. High-resolution images were taken for subsequent analysis.

**5.10 | Effect of cRGDKYQ on the viability of B16 melanoma cells**

To examine the effect of cRGDKYQ on the viability of B16 melanoma cells, log-phase B16 cells were utilized. B16 cells were seeded at a density of $5 \times 10^4$ cells per well in 24-well plates, with 500 μL of complete DMEM medium (containing 10% fetal bovine serum) added to each well. The cells were incubated at 37 °C with 5% $CO_2$ for 24 hours to allow for attachment and growth. cRGDKYQ was dissolved in sterile physiological saline to prepare solutions of varying concentrations (0, 40, 80, 200 μg/mL). After the initial 24-hour culture period, the cells in the 24-well plates were treated with these different concentrations of cRGDKYQ solution for an additional 24 hours. The control group received only sterile physiological saline without cRGDKYQ. Following the 24-hour treatment, cells were gently washed twice with PBS to remove residual cRGDKYQ solution. Calcein AM was diluted in serum-free DMEM to a final concentration of 2 μM, and propidium iodide (PI) was diluted to a final concentration of 1 μg/mL. Each well received 200 μL of this staining solution and was incubated in the dark for 30 minutes. Cells were then washed twice with PBS. Fluorescence images were captured using a confocal laser scanning microscope (Nikon A1, Japan), with appropriate excitation and emission wavelengths: Calcein AM (excitation 488 nm, emission 515 nm) for live cells and PI (excitation 535 nm, emission 617 nm) for dead cells. Images were taken under identical conditions for all groups. Green fluorescence indicated live cells stained by Calcein AM, while red fluorescence represented dead cells stained by PI.

**5.11 | In vivo efficacy validation experiment**

To evaluate the inhibitory effects of a cyclic peptide drug on B16 melanoma cells in nude mice, the following standardized protocol was employed: B16 melanoma cells were used. Cells were dissociated using trypsin and resuspended to form a single-cell suspension. Cell concentration was adjusted to $1 \times 10^6$ cells/mL using a hemocytometer for accurate counting. Cells were then re-suspended in serum-free DMEM medium (Gibco) to achieve a final concentration of $1 \times 10^6$ cells/200 μL. A volume of 200 μL cell suspension was inoculated subcutaneously into the right dorsal side of 7-week-old female nude mice. Each group consisted of 5 nude mice, including a control group and an experimental group. Post-inoculation, the body weight and tumor formation of the nude mice were monitored daily, and tumor growth was recorded. Noticeable tumor formation was observed within 7-10 days post-inoculation. Tumor dimensions were measured using calipers, recording the longest diameter (a) and shortest diameter (b), and calculating tumor volume using the formula $V = ab^2/2$. The cyclic peptide and its non-cyclic counterpart were dissolved in sterile physiological saline and administered at a concentration of 7.5 mg/kg. The control group received an equivalent volume of sterile physiological saline. Starting from day 7 after visible tumor formation, injections were administered every three days for a total of four injections. Before and after each injection, the body weight and tumor volume of the nude mice were recorded to assess the drug's effect on tumor growth inhibition. Survival times were also documented to evaluate the impact of the drug on survival rates. Seven days after the last injection, the nude mice were euthanized by cervical dislocation, and tumor tissues were excised, weighed, and prepared for further histological and molecular biological analysis. Data analysis was performed using Origin software. Comparisons between the control and experimental groups were made regarding tumor volume, changes in body weight, and survival rates. Statistical analyses were conducted using T-tests and ANOVA to determine significant differences.

**5.12 | Morphological observation of lung and tumor tissues**

To evaluate the inhibitory effects of the cyclic peptide drug on B16 melanoma cells in nude mice and its impact on lung tissue, the following protocol was employed for morphological assessment: Upon euthanizing the nude mice, tumor and lung tissues were promptly excised and gently rinsed with PBS. All tissues were fixed in paraformaldehyde, embedded in paraffin, sectioned (Leica RM2255, Germany), and stained with hematoxylin and eosin (H&E). Finally, the HE-stained sections were examined and imaged using an optical microscope (Pannoramic MIDI, Hungary). Histopathological evaluation of the tumor and lung tissue sections focused on cellular morphology, tissue architecture, and inflammatory responses, providing detailed descriptions of these features.

**Author Contributions**

R.Y.Z and J.G. designed the experiments wrote the original draft. R.Y.Z., J.G., X.W.W., Q.Q.S. and X.M.Z conducted peptide characterization. R.Y.Z. and J.J.H. conducted the animal and cell experiments. F.Z. and .M W. were the administrators of the project and participated in writing-reviewing, and editing and supervision. All the authors had read and approved the final version of the manuscript.

**Acknowledgments**

This study was funded by the National Natural Science Foundation of China (No. 82201594, T2241002, 32271298, 11975297), National Key Research Project of MOST (No. 2023YFA0915000), the opening grants of Innovation laboratory of Terahertz Biophysics (No. 23-163-00-GZ-001-001-02-01), the National Key Research and Development Program of China (No.2021YFA1200402) and Wenzhou Institute of the University of Chinese Academy of Sciences (WIUCASQD2021003, WIUCASQD2023012).

**Ethics Statement**

All animal procedures were authorized by the Animal Ethics Committee of the Wenzhou Institute, University of Chinese Academy of Sciences (Wenzhou Institute of Biomaterials & Engineering) (WIUCAS24082904)

**Declaration of competing interest**

The authors declare that they have no known competing financial interests or personal relationships that could have appeared to influence the work reported in this paper.

**Data availability**

Data will be made available on request.

**References:**

1. Faries MB, Thompson JF, Cochran AJ, et al, "Completion Dissection or Observation for Sentinel-Node Metastasis in Melanoma," *N Engl J Med* 376, no.23 (2017): 2211-2222.
2. Siegel RL, Miller KD, Wagle NS, Jemal A, "Cancer statistics, 2023," *CA: A Cancer Journal for Clinicians* 73, no.1 (2023): 17-48.
3. Knight A, Karapetyan L, Kirkwood JM, "Immunotherapy in Melanoma: Recent Advances and Future Directions," *Cancers* 15, no.4 (2023): 1106.
4. Logesh R, Prasad SR, Chipurupalli S, Robinson N, Mohankumar SK, "Natural tyrosinase enzyme inhibitors: A path from melanin to melanoma and its reported pharmacological activities," *Biochimica et Biophysica Acta (BBA) - Reviews on Cancer* 1878, no.6 (2023): 188968.
5. Liu Y, Zhao H, Li L, et al, "A tyrosinase-activated Pt(II) complex for melanoma photodynamic therapy and fluorescence imaging," *Sensors and Actuators B: Chemical* 374, (2023): 132836.
6. Xia M, Wang Q, Liu Y, et al, "Self-propelled assembly of nanoparticles with self-catalytic regulation for tumour-specific imaging and therapy," *Nature Communications* 15, no.1 (2024): 460.
7. Huang Y, Shi Y, Wang Q, et al, "Enzyme responsiveness enhances the specificity and effectiveness of nanoparticles for the treatment of B16F10 melanoma," *Journal of Controlled Release* 316, (2019): 208-222.
8. Vorobjev I, "Editorial: Cytoskeleton in the focus of anti-cancer therapy: In a search of novel biomarkers and combinatorial therapy approaches," *Frontiers in Pharmacology* 13, (2022):
9. Lackner LL, Horner JS, Nunnari J, "Mechanistic Analysis of a Dynamin Effector," *Science* 325, no.5942 (2009): 874-877.
10. Liang T, Lu L, Song X, Qi J, Wang J, "Combination of microtubule targeting agents with other antineoplastics for cancer treatment," *Biochimica et Biophysica Acta (BBA) - Reviews on Cancer* 1877, no.5 (2022): 188777.
11. Mukhtar E, Adhami VM, Mukhtar H, "Targeting Microtubules by Natural Agents for Cancer Therapy," *Molecular Cancer Therapeutics* 13, no.2 (2014): 275-284.
12. Pierschbacher MD, Ruoslahti E, "Cell attachment activity of fibronectin can be duplicated by small synthetic fragments of the molecule," *Nature* 309, no.5963 (1984): 30-33.
13. Ruoslahti E, "RGD and other recognition sequences for integrins," *Annu Rev Cell Dev Biol* 12, 697-715.
14. Yin L, Li X, Wang R, Zeng Y, Zeng Z, Xie T, "Recent Research Progress of RGD Peptide‐Modified Nanodrug Delivery Systems in Tumor Therapy," *International Journal of Peptide Research and Therapeutics* 29, no.4 (2023): 53.
15. Qin W, Chandra J, Abourehab MAS, et al, "New opportunities for RGD-engineered metal nanoparticles in cancer," *Molecular Cancer* 22, no.1 (2023): 87.

16. Amin M, Mansourian M, Koning GA, Badiee A, Jaafari MR, ten Hagen TLM, "Development of a novel cyclic RGD peptide for multiple targeting approaches of liposomes to tumor region," *Journal of Controlled Release* 220, (2015): 308-315.
17. Sanati M, Afshari AR, Aminyavari S, Kesharwani P, Jamialahmadi T, Sahebkar A, "RGD-engineered nanoparticles as an innovative drug delivery system in cancer therapy," *Journal of Drug Delivery Science and Technology* 84, (2023): 104562.
18. Guo J, Zhao R, Gao F, Li X, Wang L, Zhang F, "Sequence-Dependent Tyrosine-Containing Peptide Nanoassemblies for Sensing Tyrosinase and Melanoma," *ACS Macro Letters* 11, no.7 (2022): 875-881.
19. Guo J, Li X, Lian J, et al, "Green Fluorescent Tripeptide Nanostructures: Synergetic Effects of Oxidation and Hierarchical Assembly," *ACS Macro Letters* 10, no.7 (2021): 825-830.
20. Guo J, Zheng F, Song B, Zhang F, "Tripeptide-dopamine fluorescent hybrids: a coassembly-inspired antioxidative strategy," *Chemical Communications* 56, no.46 (2020): 6301-6304.
21. Guo J, Ramachandran S, Zhong R, Lal R, Zhang F, "Generating Cyan Fluorescence with De Novo Tripeptides: An In Vitro Mutation Study on the Role of Single Amino Acid Residues and Their Sequence," *ChemBioChem* 20, no.18 (2019): 2324-2330.
22. Park J, Wu Y, Suk Kim J, Byun J, Lee J, Oh Y-K, "Cytoskeleton-modulating nanomaterials and their therapeutic potentials," *Advanced Drug Delivery Reviews* 211, (2024): 115362.
23. Sun M, Wang C, Lv M, Fan Z, Du J, "Intracellular Self-Assembly of Peptides to Induce Apoptosis against Drug-Resistant Melanoma," *Journal of the American Chemical Society* 144, no.16 (2022): 7337-7345.
24. Tang M, Duan T, Lu Y, Liu J, Gao C, Wang R, "Tyrosinase-Woven Melanin Nets for Melanoma Therapy through Targeted Mitochondrial Tethering and Enhanced Photothermal Treatment," *Advanced Materials* 36, no.44 (2024): 2411906.